\documentclass[aps,prl,twocolumn,showpacs]{revtex4}
\usepackage{bm,bbm,amsmath, amssymb,amsbsy,graphicx,amsthm,times,txfonts}
\usepackage{hyperref}
\usepackage{color}
 \newcommand{\tr}[1]{\text{Tr}}

\begin{document}

\title{Observable Measure of Bipartite Quantum Correlations}

\author{Davide Girolami}
\email{pmxdg1@nottingham.ac.uk}
\author{Gerardo Adesso}
\email{gerardo.adesso@nottingham.ac.uk}
\affiliation{$\mbox{School of Mathematical Sciences, University of Nottingham, University Park, Nottingham NG7 2RD, United Kingdom}$}

\date{April 6, 2012}

\begin{abstract}
We introduce a measure $Q$ of  bipartite quantum correlations for arbitrary two-qubit states, expressed as a state-independent function of the density matrix elements. The amount of quantum correlations can be quantified experimentally by measuring the expectation value of a small set of observables on up to four copies of the state, without the need for a full tomography. We extend the measure to  $2\times d$ systems, providing its explicit form in terms of observables and applying it to the relevant class of multiqubit states employed in the  deterministic quantum computation with one quantum bit model. The number of required measurements to determine $Q$ in our scheme does not increase with $d$. Our results provide an experimentally friendly framework to estimate quantitatively the degree of general quantum correlations in composite systems.
 \end{abstract}


\maketitle

Quantum entanglement is one of the most fundamental consequences of the superposition principle and undoubtedly  plays a key role in designing faster-than-classical algorithms, teleportation protocols and super-dense coding \cite{entanglement}. However, it may not be the ultimate resource behind the power of quantum computation \cite{merali}. It has been recently found that, even with no entanglement,  some mixed-state based schemes [such as the so-called deterministic quan-
tum computation with one quantum bit (DQC1) \cite{laf1}] allow an improvement of performance in computing tasks  \cite{dattabarbieriaustinchaves}, and, more generally, separable states possess genuinely quantum correlations \cite{modirev} (QCs), captured e.g.~by the quantum discord \cite{OZ,HV}, which cannot be described within a classical scenario \cite{OZ, HV, nonlocwithoutent, req, piani, MID}.
In general, QCs in a state $\rho_{AB}$ can be defined as the minimum amount of total correlations (measured, e.g., by the mutual information) between Alice and Bob, that are destroyed by a local measurement on one or both subsystems \cite{OZ,HV,MID,altremisure,req,modi,dakic,genio}. For pure states, QCs coincide with entanglement \cite{OZ}. For mixed states, even if some operational interpretations have been proposed \cite{piani, dattamerging, luofu, streltsov,genio}, basic technical issues still prevent us from reaching a full comprehension of their nature. Indeed, theoretical evaluation and experimental detection of QCs both represent hard challenges: any attempt to determine the QCs in a given state $\rho_{AB}$ is hindered by the difficulty of solving an optimisation to determine the least disturbing measurement for that state, thus requiring the full knowledge of it. Recently, some non-tomographic detection schemes for witnessing non-vanishing QCs by measuring just one observable \cite{dakic,mazi, cinesiwitness} have been proposed and implemented. However, noting that {\it all} states possess nonzero QCs but a null-measure set \cite{acinferraro}, the most worthwhile question becomes that of evaluating, by a proper measure, the actual {\it amount} of QCs encoded in a state. Only then, quantitative connections can be drawn between the QCs content and the performance of some quantum protocol using them as resource \cite{dattabarbieriaustinchaves,chaves,remote}.

In this Letter, we show that QCs in a general two-qubit state $\rho$ can be reliably quantified without any explicit optimisation and with no need to know the full shape of the state. We define a QCs measure $Q$ which is a state-independent function of the density matrix elements. In particular, $Q$ can be expressed in terms of the expectation values of a set of nine observables $\{O_i\}$. Consequently, such a function could be evaluated by designing simple quantum circuits simulating the measurements of $\{\langle O_i\rangle\}$ \cite{paisa12, paz, brun, ekert, filip1,filip2, winter, horo0}. However, following the alternative approach of Refs.~\cite{mint1,mintexp,cina}, we further show that the quantity $Q$ can be even less demandingly measured by performing seven local projections on up to four copies of the state $\rho$ .
Then, we extend our measure to capture bipartite QCs in states of $2 \times d$ dimensional systems, finding that seven projective measurements are always sufficient to experimentally determine $Q$, i.e.~the number of measurements required is independent of  $d$. Specifically, we use this construction to obtain a quantitative estimate of QCs in a recent experimental implementation \cite{laf2} of the DQC1 model \cite{laf1} with four qubits.

\begin{figure}[tb]
\includegraphics[width=7cm]{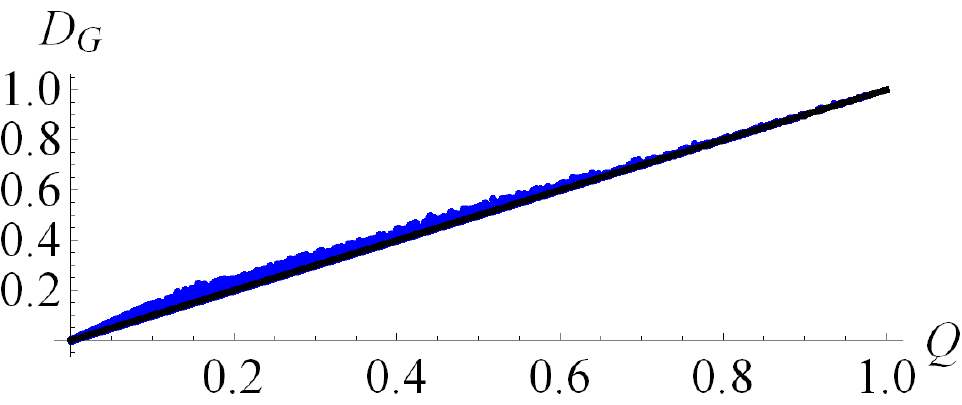}
  \caption{Geometric discord $D_G$ versus its  tight lower bound $Q$ for $3 \cdot 10^4$ random two-qubit states. The plotted quantities are dimensionless.}
    \label{geo}
\end{figure}

A number of conceptually different measures of general QCs have been recently proposed \cite{modirev,OZ,HV,MID,altremisure,req,modi,dakic,genio}. In the following, we consider a two-qubit state $\rho\equiv\rho_{AB}$ and adopt a geometric perspective, quantifying the QCs in terms of the minimum distance of $\rho$ from the set $\Omega$ of classical-quantum states. The states $\chi \in \Omega$ filling such set are left unperturbed by at least one choice of projective measurement on Alice, and take the form \cite{piani} $\chi= \sum_i p_i |i\rangle\langle i| \otimes \rho_{i B} $, where $p_i$ are probabilities, $\{|i\rangle\}$ is an orthonormal vector set and $\rho_{i B}$ is the marginal density matrix of Bob. Adopting the Hilbert-Schmidt norm   $\|M\|_2=\sqrt{\text{Tr}(M M^\dagger)}$, one obtains a QCs measure known as ``geometric discord'', introduced in \cite{dakic}, operationally interpreted in \cite{luofu,remote}, and defined as
\begin{equation}D_G(\rho)=2 \min_{\chi \in \Omega}  \|\rho -\chi \|_2^2\,,
\end{equation}
where we add a normalisation factor $2$.  The geometric discord enjoys a closed expression for two-qubit states. First, one needs to express the state in the Bloch basis,
 $   \rho = \frac14 \sum_{i,j=0}^3 R_{ij} \sigma_i \otimes \sigma_j = \frac 14(\mathbb{I}_{4}+\sum_{i=1}^3 x_i\sigma_i \otimes \mathbb{I}_{2} +\sum_{j=1}^3 y_j \mathbb{I}_{2}\otimes \sigma_j+\sum_{i,j=1}^3 t_{ij} \sigma _i\otimes\sigma_j)$,
where $R_{ij}=\text{Tr}[\rho(\sigma_i\otimes \sigma_j)]$, $\sigma_0=\mathbb{I}_{2}$, $\sigma _i$ ($i=1,2,3$) are the Pauli matrices, $\vec{x}=\{x_i\},\vec{y}=\{y_i\}$ represent the three-dimensional Bloch column vectors associated to $A,B$, and  $t_{ij}$ are the elements of the correlation matrix $t$.  Then, following \cite{dakic}, we have:
 $D_G(\rho)= \frac 12(\|\vec x\|^2 + \|t\|_2^2 -4 k_{\max})=2 \text{Tr}[S]-2k_{\max}$,
  with $k_{\max}$ being the largest eigenvalue of the matrix $S = \frac 14(\vec x {\vec  x}^{\sf T}+  t t^{\sf T})$. We now provide an explicit expression for $k_{\max}$.  The characteristic equation of the matrix $S$ is cubic and can be solved analytically \cite{cubica}. Being constrained to real solutions only, we write the eigenvalues $\{k_i\}$ of $S$ as
\begin{eqnarray}\label{k}
k_i=\frac{\text{Tr}[S]}3+\frac{\sqrt{6\text{Tr}[S^2]-2\text{Tr}[S]^2}}3\cos\left(\frac{\theta+  \alpha _i}{3}\right),
\end{eqnarray}
where $\{\alpha_i\}=\{0, 2\pi,4\pi\}$ and $\theta = \arccos\big\{(2 \text{Tr}[S]^3-9 \text{Tr}[S]\text{Tr}[S^2]+9\text{Tr}[S^3])
\sqrt{2/(3\text{Tr}[S^2]-\text{Tr}[S]^2)^3}\big\}$. Since $\theta$ is an arccosine, we have  $0 \leq  \theta/3\leq \pi /3$ and the maximum of $\cos{\frac{\theta+\alpha_i}{3}}$ is reached for $\alpha_i\equiv\alpha_1=0$. Hence,  $k_{\max}\equiv\text{max}\{k_i\}= k_1$, and the geometric discord for an arbitrary two-qubit state $\rho$ assumes the form of a state-independent function of its entries $(\rho_{ij})$, that is
\begin{eqnarray}\label{dg}
D_G(\rho)=2 (\text{Tr}[S]-k_1).
\end{eqnarray}
However, we aim to define a simpler, and more  accessible experimentally, QCs quantifier. By replacing $\theta$ with $0$ in Eq.~\eqref{k}, we obtain a meaningful and remarkably tight lower bound $Q \le D_G$ (see Fig.~\ref{geo}) to the geometric discord, given by
\begin{eqnarray}\label{q}
Q(\rho)=\mbox{$\frac23$}\left( {2\text{Tr}[S]}-{\sqrt{6\text{Tr}[S^2]-2\text{Tr}[S]^2}} \right).
\end{eqnarray}
 The quantity $Q$ is still a state-independent expression of the entries of $\rho$, but a rather easier one to manage than $D_G$, and can be regarded as a {\it bona fide} measure of QCs in its own right. Indeed, it is non-negative by definition, it is faithful (i.e., vanishes only on classical-quantum states $\chi$) and coincides with $D_G$ for pure states. The latter two properties can be proven as follows. Faithfulness is equivalent to showing that $Q=0 \iff D_G =0$;  the condition for vanishing $Q$ is $\text{Tr}[S]^2=\text{Tr}[S^2]$, and by the Cayley-Hamilton theorem this implies $\text{Tr}[S]^3=\text{Tr}[S^3]$, i.e.,  $D_G=0$. The equality between $D_G$ and $Q$ for pure states follows from the fact that, writing a bipartite pure state (where subsystem $A$ is a qubit) in the Schmidt decomposition, we have $\theta=0$.
  We also find that  $Q$ provides a non-trivial upper bound on an entanglement measure, specifically the squared negativity ${\cal N}^2$ \cite{entanglement}. In fact,  the chain $D_G \geq Q \geq {\cal N}^2$ holds for arbitrary two-qubit states, with all inequalities saturated on pure states \cite{xfile}.  Finally, let us mention that a simple upper bound on $D_G$ can be obtained as well from Eq.~\eqref{k}, $D_G \le 4\text{Tr}[S]/3$.

From now on, we adopt $Q$ as a rightful QCs quantifier for two qubits, endowed with the advantage of requiring neither theoretical optimisations nor experimental state tomography for its evaluation.  Specifically, the task of providing a recipe for measuring $Q$  reduces to write $\text{Tr}[S]$ and  $\text{Tr}[S^2]$ as functions of suitable observables, and is accomplishable as follows. Defining the matrices  $X=\vec{x}{\vec{x}}^{\sf T}, T=t t^{\sf T}$, we have $\text{Tr}[S]=(\text{Tr}[X]+\text{Tr}[T])/4 =\text{Tr}[ \rho^2]-\text{Tr}[ \rho_B^2]/2$,
 and  $\text{Tr}[S^2]=\frac 1{16}(\text{Tr}[X^2]+\text{Tr}[T^2]+2 \text{Tr}[XT])$.
 After some algebra we obtain
\begin{eqnarray}\label{sigma}
\text{Tr}[S^2]=&\frac 14&(-2-8\text{Tr}[\rho^4]+8\text{Tr}[\rho^3]+ 6\text{Tr}[\rho^2]^2\\
&-&2 \text{Tr}[\rho^2](5+\text{Tr}[\rho_B^2])-2 \text{Tr}[\rho_A^2]^2+10\text{Tr}[\rho_A^2]\nonumber\\
&-&\text{Tr}[\rho_B^2]^2+12\text{Tr}[\rho_B^2]-6\text{Tr}[\rho_A^2]\text{Tr}[\rho_B^2]\nonumber\\
&+&4\text{Tr}[\rho (\mathbb{I}_2\otimes\rho_B)\rho(\mathbb{I}_2\otimes\rho_B)]-24\text{Tr}[\rho (\rho_A\otimes\rho_B)]\nonumber\\
&+&8\text{Tr}[\rho (\rho_A\otimes\mathbb{I}_2)\rho(\rho_A\otimes\mathbb{I}_2)]+8\text{Tr}[\rho^2(\rho_A\otimes\rho_B)])\nonumber .
\end{eqnarray}
By substituting  in Eq.~\eqref{q}, $Q$ takes the form of a function of fourth-order polynomials of $(\rho_{ij})$, in particular it is written in terms of traces of matrices powers. Now, given a general density matrix $\rho$, it holds \cite{paisa12, paz,brun, ekert, winter, horo0}  $\text{Tr}[\rho^k]=\text{Tr}[V^k \rho^{\otimes k}]$, where $V^k$ is the shift operator, $V^k|\psi_1\psi_2\ldots\psi_k\rangle=|\psi_k\psi_1\ldots\psi_{k-1}\rangle$. Also, for two unknown states $\rho_1,\rho_2$, it has been proven \cite{ekert} that $ \text{Tr}[V^2 \rho_1\otimes \rho_2]=\text{Tr}[\rho_1\rho_2]$.  More generally, we have \cite{linden} $
\text{Tr}[\rho_1\rho_2\ldots\rho_k]=\text{Tr}[V^k \rho_1\otimes \rho_2\otimes\ldots\otimes\rho_k]$.

\begin{figure}[tb]
 \includegraphics[width=8.5cm]{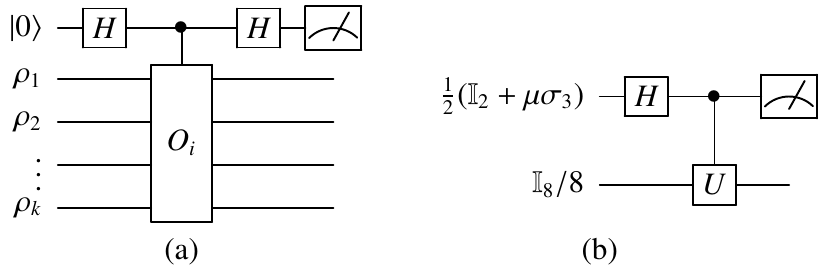}
  \caption{(a) Quantum circuit estimating $\text{Tr}[O_i \rho_1\otimes \rho_2\otimes\ldots\otimes\rho_k]= v$.  Two Hadamard gates $H$ are applied to an ancilla, followed by a measurement in the computational basis. The interferometer is modified by inserting a controlled-$O_i$ gate acting on the overlap of states. \\ (b) DQC1 model with a register of three maximally mixed qubits and an ancillary qubit in a state of polarisation $\mu$. Expectation values of $\sigma_1, \sigma_2$ on the ancilla return the real and imaginary part of $\text{Tr}[U]/8$.}
    \label{magic}
\end{figure}

We can exploit these results and follow the approach proposed in \cite{ekert} for estimating the expectation values of the appropriate unitary operators $\{O_i\}_{i=1}^9$ to associate with each of the nine independent factors in Eq.~\eqref{sigma} (which include those appearing in $\text{Tr}[S]$ as well). They can all be expressed as shift operators $O_i=V^k$ on a number $k$ ($k \le 4$) of copies of the global and/or marginal density matrices and their overlaps, depending on each particular term in Eq.~\eqref{sigma}.
The circuit to be implemented, which includes an ancillary meter qubit, is depicted in  Fig.\ref{magic}(a).  For each $O_i$, we build an interferometer modified by inserting a controlled-$O_i$ gate: defining the visibility $v$,  we obtain in general $\text{Tr}[O_i \rho_1\otimes \rho_2\otimes\ldots\otimes\rho_k]= v$. Hence, QCs can be measured quantitatively  from the expectation values of the nine operators $\{O_i\}$ only---as opposed to $15$ observables required for complete state tomography---obtained in laboratory via  readouts on the ancillary qubit.

We wish now to provide an alternative scheme for the exact measurement of $Q$ that further reduces the required resources for its implementation, and appears even more experimentally friendly. This is done by rephrasing the detection scheme in terms of  local (with respect to the Alice-Bob split) projectors on multiple (up to $4$) copies $\rho^{\otimes n}$ of the {\it same} state $\rho$ \cite{filip1,filip2,mint1,mintexp,cina}.
We observe that
\begin{eqnarray}\label{eq0}
\text{Tr}[S]=\text{Tr}[ \rho^2]-\text{Tr}[ \rho_B^2]/2\,.
\end{eqnarray}
It is known \cite{filip2} that $\text{Tr}[\rho^2]=\text{Tr}[V^2\rho^{\otimes 2}]=\text{Tr}[(P^+-P^-)\rho^{\otimes 2}]=1-2\text{Tr}[P^-\rho^{\otimes 2}]$,
where $V^2$ is the swap operator and $P^{\pm}$ are the projectors on the symmetric/antisymmetric subspaces. Named $A_i$ ($B_j$)  the subsystems controlled by Alice (Bob) in the $i$-th ($j$-th) copy of the bipartite state $\rho$, we have then
\begin{equation}\label{eq1}
\text{Tr}[S]=\frac12- 2 \text{Tr}[ P_{(A_1B_1)(A_2B_2)}^-\rho^{\otimes 2}] +  \text{Tr}[ P_{B_1B_2}^-\rho_B^{\otimes 2}]\,,
\end{equation}
where for two qubits $P_{B_1B_2}^-=|\psi_{B_1B_2}^{-}\rangle\langle \psi_{B_1B_2}^{-}|, |\psi_{B_1B_2}^{-}\rangle=\frac1{\sqrt{2}}(|01\rangle-|10\rangle)$, while
$P_{(A_1B_1)(A_2B_2)}^-=\frac1{8}\left(3\mathbbm{I}_{16} - \sum_i \sigma^{(4)}_i \otimes \sigma^{(4)}_i \right)$, where with $\sigma^{(d)}$ we indicate the generalised (and normalised) Gell-Mann matrices for dimension $d$.
 Alternatively, we can exploit the very recent  results of Ref.~\cite{cina} and write
 \begin{eqnarray}\label{eq2}
 \text{Tr}[S]&=&4 c_1 - 2 c_2 -  c_3 +\mbox{$\frac12$}\,, \nonumber\\
 c_1&=&\text{Tr}[(P^-_{A_1A_2}\otimes P^-_{B_1B_2}) \rho^{\otimes 2}]\,;\nonumber\\
 c_2&=&\text{Tr}[(P^-_{A_1A_2}\otimes \mathbb{I}_{B_1B_2}) \rho^{\otimes 2}]\,;\nonumber\\
 c_3&=&\text{Tr}[(\mathbb{I}_{A_1A_2}\otimes P^-_{B_1B_2}) \rho^{\otimes 2}]\,.
 \end{eqnarray}
In \cite{filip2} a method to measure the purity of a quantum state, which is all that we need, is presented and demonstrated by means of the implementation of an all-optical set up. Ref.~\cite{cinaexp} presents a more comprehensive detection scheme for projective measurements; see also \cite{pascazio} for a very recent alternative method. To sum up, in this framework we need three measurements of two-qubit projectors [Eq.~(\ref{eq2})] ---or two measurements, one on two qubits, the other (non-local with respect to the Alice-Bob split) on four qubits [Eq.~(\ref{eq1})]--- and two copies of the state, to measure ${\text{Tr}[S]}$.

The detection of $\text{Tr}[S^2]$ can also be recast in terms of local projections. Following Ref.~\cite{cina}, we obtain
    \begin{eqnarray}\label{eq3}
    \text{Tr}[S^2]&=&16 c_4 + 8 (c_7 - c_5 - 2 c_6) +c_3^2 + 4c_2^2 - c_3 - 2c_2\
+ \mbox{$\frac14$}\,,\nonumber\\
    c_4&=&\text{Tr}[(P^-_{A_1A_4}\otimes P^-_{A_2A_3}\otimes P^-_{B_1B_2}\otimes P^-_{B_3B_4}) \rho^{\otimes 4}]\,;\nonumber\\
    c_5&=&\text{Tr}[(P^-_{A_1A_4}\otimes \mathbb{I}_{A_2A_3} \otimes P^-_{B_1B_2}\otimes P^-_{B_3B_4}) \rho^{\otimes 4}]\,;\nonumber\\
    c_6&=&\text{Tr}[(P^-_{A_1A_4}\otimes P^-_{A_2A_3} \otimes P^-_{B_1B_2}\otimes \mathbb{I}_{B_3B_4}) \rho^{\otimes 4}]\,;\nonumber\\
    c_7&=&\text{Tr}[(\mathbb{I}_{A_1A_4}\otimes P^-_{A_2A_3} \otimes P^-_{B_1B_2}\otimes  \mathbb{I}_{B_3B_4}) \rho^{\otimes 4}].
    \end{eqnarray}
Compared to the measurement of $\text{Tr}[S]$, here we have again projectors on pairs of qubits, however they need to be implemented on four copies of the state $\rho$. As Eq.~(\ref{eq3}) shows, we can evaluate the value of $\text{Tr}[S^2]$ by measuring four such independent projectors in laboratory.
  Therefore, in the most economical scheme devised here, \emph{the full quantitative detection of bipartite QCs in an arbitrary two-qubit state $\rho$ as measured by $Q$ demands six or seven projective measurements on (up to) four copies of the state $\rho$.} Notice for comparison that to measure the geometric discord $D_G$ exactly, Eq.~(\ref{dg}), one would need $11$ projective measurements on up to six copies of the state \cite{cina}. On the other hand, at a qualitative level, a single observable witness suffices to reveal whether $Q$ (or the discord) is zero or not \cite{dakic,laf2,mazi,cinesiwitness}. 

 We now extend our measure to higher-dimensional systems, in particular to $2 \times d$ systems, which include the practically relevant case of one qubit ($A$) versus a register ($B$) of $n$ qubits. The geometric discord for an arbitrary state $\rho$ of a $2\times d$ system has been derived in \cite{rau}, and has the same expression as Eq.~\eqref{dg}, just amended with the following generalisations: $\text{Tr}[S]=\frac{1}{2d}(\text{Tr}[X]+\text{Tr}[T])$,  $x_i=\text{Tr}[\rho( \sigma_i\otimes\mathbb{I}_d)]$, and $t_{ij}=\text{Tr}[\rho (\sigma_i\otimes \tau_j)]$,  where, for example, we can assume  $\{\tau_j\}\equiv\{\sigma^{(d)}_j\}$ as $d$-dimensional basis for Bob's subsystem.
  We can repeat the steps done for $2 \times 2$ systems and obtain a state-independent form for $D_G$ and for the lower bound $Q$ as well, since the subsystem $A$ is still a qubit and therefore the characteristic equation of $S$ remains a cubic.
 Thus, for $2\times d$ states, the task we face is again to express $\text{Tr}[S], \text{Tr}[S^2]$ in terms of observables.

 The most practical way to proceed is to consider the scheme in terms of local projectors. In this respect, it is straightforward to verify that Eqs.~(\ref{eq1}, \ref{eq2}, \ref{eq3}) still hold for $2\times d$ systems: their expression can be written in exactly the same form as for the two-qubit case, provided we generalise the swap and the projectors $P^-$ to arbitrary dimension $d$ as follows, $V^2=\frac1d(\mathbb{I}_{d^2}+\sum_i \tau_i\otimes\tau_i)$
and $P^-_{S_j S_k}=\frac1{2d}((d-1)\mathbb{I}_{d^2}-\sum_i \tau_i\otimes\tau_i)$, where $S_j, S_k$ denote two $d$-dimensional systems, and the $\tau_i$'s reduce to Pauli matrices in dimension $d=2$ (e.g., when we want to calculate $\text{Tr}[\rho_B^2]$ in the $2 \times 2$ case).  This observation, combined with the previous analysis, allows us to conclude that, even \emph{for arbitrary states $\rho$ of ${2\times d}$ dimensional systems, we just need six or seven projective measurements on up to four copies of the state $\rho$ to quantify bipartite QCs between the qubit and the remaining qudit system}. The number of measurement settings thus does not increase with $d$, which demonstrates the efficiency and scalability of our scheme. Clearly, the optical implementation of projectors of the type $P_{B_iB_j}$, i.e.~multi-qubit projectors, is more complicated than the two-qubit case, see e.g.~\cite{ent}.  However, the method demonstrated in \cite{cinaexp} can be extended to arbitrary dimensions without dramatically increasing the complexity of the experimental setting (as claimed by the authors in the last section of Ref.~\cite{cinaexp}). More precisely, the number of optical elements required to implement each projector (basically interferometers)  should increase polynomially---namely, linearly---with $d$ \cite{cinaexp,2cinaexp}, in stark contrast with a complete quantum state tomography for which the required resources scale exponentially \cite{jensnat}.
 Note also that our scheme  for $2 \times d$ systems is completely general and no prior knowledge of the form of the state is required; it only relies on the implicit assumption that the subsystem $A$ has dimension $2$, i.e., it is indeed a qubit. This assumption can be verified in laboratory {\it a priori} e.g.~by measuring suitable Hilbert space dimension witnesses \cite{dimwitness}, or, possibly, with tomography on the marginal state of subsystem $A$, which only consumes a fixed, small amount of extra resources.

\begin{figure}[tb]
\includegraphics[width=8cm]{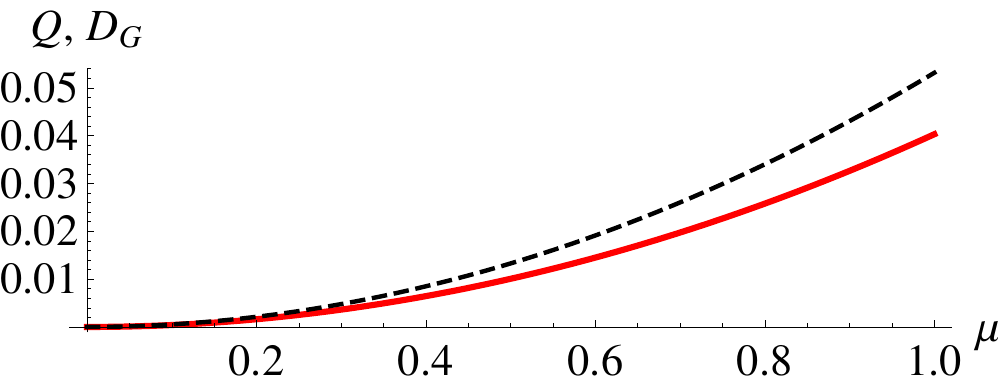}
  \caption{Bipartite quantum correlations as measured by $Q$ (red continuous line)  and $D_G$ (black dashed line) for the output state of the DQC1 model as implemented in Ref.~\cite{laf2} with four qubits, plotted as functions of the initial polarisation $\mu$ of the ancilla qubit. All the plotted quantities are dimensionless.}
    \label{QvsD}
\end{figure}

  We consider as an example the four-qubit implementation of the DQC1 model for quantum computation \cite{laf1}.
  This algorithm estimates the trace of a normalised unitary matrix $U$.  We consider the instance recently implemented experimentally in \cite{laf2} (where only a discord witness rather than a quantitative estimate was measured), where $U=(a,a,b,1,a,b,1,1)$, with $a=-(e^{-i 3\pi/5})^4, b=(e^{-i 3\pi/5})^8$. Such a specific gate is used for the approximation of Jones polynomials \cite{jones}.   The first qubit (the ancilla) is initially in a state of polarisation $\mu$, while the remaining qubits are maximally mixed. Referring to the scheme of Fig.~\ref{magic}(b),
    the final state of the system before readout is
  \begin{eqnarray}
\rho_{out}=\frac1{16}\left(
\begin{array}{c|c}
 \mathbb{I}_{8} & \mu U^{\dagger}  \\ \hline
 \mu U &   \mathbb{I}_8 \\
\end{array}
\right).
  \end{eqnarray}
 We calculate the bipartite QCs $Q(\rho_{out})$, measurable in laboratory according to the scheme detailed above, between the ancilla and the residual three-qubit system, and we compare them with the geometric discord $D_G$, while entanglement is always zero across this bipartition. The plots in Fig. \ref{QvsD}  reveal that $Q$ is in good agreement with $D_G$, being a monotonic function of the polarisation $\mu$ of the ancilla, hence showcasing its reliability as a QCs quantifier \cite{dattabarbieriaustinchaves}.

In conclusion, we presented a scheme to quantify theoretically and experimentally general bipartite QCs for arbitrary two-qubit and qubit-qudit states. We introduced a measure $Q$ that is a state-independent function of polynomials of the density matrix elements, and can be measured by implementing a restricted number of quantum circuits, or alternatively a restricted number of local projections, on up to four copies of the state, which appears in reach of current technology \cite{mintexp,cinaexp,2cinaexp}. We used our measure to evaluate quantitatively the degree of QCs created in a recent experimental implementation \cite{laf2} of the DQC1  model with four qubits \cite{laf1}.

Providing experimentally friendly recipes for the measure of QCs in $n$-partite realisations of quantum information protocols
is key to clarifying their usefulness for the performance of such practical tasks \cite{merali}.
In this respect, much attention is being devoted to the QCs dynamics in open quantum systems \cite{cinesinatcomm, mazzola, fanchini, cinamarkov} and, independently, to characterising the transition from Markovian to non-Markovian regimes \cite{cubitt,breuerthe, breuer3,plenio,breuerexp}.  Non-Markovianity
can be witnessed by monitoring entanglement between one subsystem, coupled to the environment, and another clean subsystem \cite{plenio}. One might imagine that more general QCs could be somehow more sensitive to the properties of dynamical maps.
However, it is known that even local Markovian channels (as well as Markovian common environments \cite{benatti}) can induce an increase of discord-like QCs in a composite system \cite{streltsov2}. Therefore, the question needs to be formulated properly and with care, and demands a dedicated analysis which is beyond the scope of this work \cite{inpr}.

We hope our Letter may contribute to render the general quantumness of correlations a more accessible (theoretically and experimentally) concept in the study of complex quantum systems.

\emph{Acknowledgments.}-- We thank the University of Nottingham for financial support through an Early Career Research and Knowledge Transfer Award and a Graduate School Travel Prize Award (Grants No. ECRKTA/2011/A2RHF4 and No. TP/SEP11/10-11/181). We warmly acknowledge discussions with B. Daki\'c, R. Filip, S. Grey, M. Guta,  P. Horodecki, R. Jozsa, S. Maniscalco, L. Mazzola, L. Mi\v{s}ta, Jr., T. Short, and V. Vedral. The circuits have been drawn by using the package \emph{Q-circuit} written by B. Eastin and S. Flammia.

\end{document}